\documentclass[twocolumn, aps, prd, superscriptaddress, showpacs, nofootinbib, floatfix, preprintnumbers]{revtex4-1}
\usepackage{epsfig,bm}
\usepackage{graphicx,amssymb}
\usepackage{amsmath}
\usepackage{epstopdf}
\usepackage[eps]{pstricks}
\usepackage[normalem]{ulem}  

\renewcommand\sout{\bgroup \color{red} \ULdepth=-.5ex \ULset}
\usepackage{slashed}
\newcommand{\comment}[1]{}

\begin{document}


\title{Case for quarkyoniclike matter from a constituent quark model}


\author{Aaron Park}%
\email{aaron.park@yonsei.ac.kr}
\affiliation{Department of Physics and Institute of Physics and Applied Physics, Yonsei University, Seoul 03722, Korea}

\author{Kie Sang Jeong}%
\email{ksjeong@uw.edu}
\affiliation{Institute for Nuclear Theory, University of Washington, Seattle, WA 98195, USA}

\author{Su Houng Lee}%
\email{suhoung@yonsei.ac.kr}
\affiliation{Department of Physics and Institute of Physics and Applied Physics, Yonsei University, Seoul 03722, Korea}

\preprint{INT-PUB-21-012}


\begin{abstract}
Based on the fact that the constituent quark model reproduces the recent lattice result on baryon-baryon repulsion at short distance and that it includes the quark dynamics with confinement, we analyze to what extent the quarkyonic modes appear in the phase space of baryons as one increases the density before  only quark dynamics and hence deconfinement occurs.
We find that as one increases the baryon density, the initial quark mode that appears will involve the $d(u)$-quark from a neutron (proton), which will leave the most attractive ($ud$) diquark intact.
\end{abstract}


\maketitle

\section{Introduction}

Recent progresses of multi-messenger astrophysics provided important clues to the properties of  dense nuclear matter~\cite{Demorest:2010bx, Antoniadis:2013pzd, Cromartie:2019kug, TheLIGOScientific:2017qsa, Abbott:2018exr}.  To support the massive  neutron stars
 whose masses are larger than two times the solar mass~\cite{Demorest:2010bx,Antoniadis:2013pzd, Cromartie:2019kug}, it was found that the equation of states (EoS) for dense matter had to be sufficiently hard.
 Meanwhile, the tidal deformability constrained via the GW170817 observation~\cite{TheLIGOScientific:2017qsa, Abbott:2018exr} from the neutron star merger implies the possible neutron star radius to be  $R_{1.4}\leq 13.5~\textrm{km}$, which requires a relatively soft EoS.  Subsequent analyses~\cite{Annala:2017llu, Raithel:2018ncd, Fujimoto:2019hxv}  showed that the EoS should have a stiff density evolution around some moderate density regime with the sound velocity $v_s^2 \geq 1/3$ to reconcile these contrary requirements.
  Furthermore, recent NICER measurement reported $R_{2.08} = 13.7+2.6-1.5~\textrm{km}$ from pulsar PSR J0740+6620~\cite{Miller:2021qha, Riley:2021pdl}, whose large mass and radius also
  supports the stiff evolution of EoS around the core density. As a possible origin for the stiff evolution, one considers an EoS built from the quarkyonic matter concept.

The quarkyonic matter concept~\cite{McLerran:2007qj} was introduced originally from the large-$N_c$ quantum chromodynamics (QCD)~\cite{tHooft:1973alw, tHooft:1974pnl}   description of cold-dense matter. If the quark Fermi momentum is large enough ($k_{F}^{q} \gg \Lambda_{\textrm{QCD}}$), the quark distributed around Fermi surface will be confined into the baryon-like state, as is the case in the vacuum, because Debye screening effect will be minimal due to the $1/N_c$ suppression of the quark loop correction to the gluon propagator. Then, the phase space distribution of the baryon-like state will be the shell-like form due to the Pauli exclusion principle between the confined quarks and the quasi-free quarks occupying the lower phase. Assuming that the quarkyonic matter configuration could appear at moderate densities, this concept has been applied to recent neutron star studies~\cite{McLerran:2018hbz, Jeong:2019lhv, Sen:2020peq, Duarte:2020xsp, Duarte:2020kvi, Zhao:2020dvu, Sen:2020qcd, Margueron:2021dtx} where it was found that the hard-soft evolution of the EoS appears naturally inside the neutron star.

The quarkyonic-like configuration with shell-like phase space distribution of baryon can be dynamically generated through the hard-core repulsive interaction between nucleons~\cite{Jeong:2019lhv, Sen:2020peq, Duarte:2020xsp, Duarte:2020kvi, Sen:2020qcd}. Although the formation of the quarkyonic-like configuration requires enhanced kinetic energy density, this configuration can exist as true ground state if the diverging repulsive baryon potential exists.  The presence of the short range repulsion in baryon-baryon interaction is an essential ingredient in providing nuclear stability~\cite{Bethe:1971xm} and our understanding of it evolved with our description of strong interaction~\cite{Lacombe:1980dr,Machleidt:1987hj,Oka:1981rj,Shimizu:1989ye,Weinberg:1990rz,Ordonez:1993tn,Bedaque:2002mn,Polinder:2007mp}.
It was recently shown that the short distance part of the baryon-baryon interaction including those with strangeness extracted from the recent lattice calculation~\cite{Ishii:2006ec,Inoue:2010hs} can be well reproduced using a constituent quark model~\cite{Park:2019bsz}.
This result together with the fact that the constituent quark model involves the quark dynamics including the confinement physics,  suggests that
by analyzing the quark dynamics in the presence of  neighbouring nucleons, one can  realistically assess if the quarkyonic modes naturally   appear as one increases the density before only quark dynamics and hence deconfinement sets in.

From a phenomenological model point of view, the quarkyonic-like configuration at finite baryon density can be understood as the appearance of a modified  phase space distribution from the ground state composed of quasi-baryons.   At shorter distance from its equilibrium position or larger momentum scale, baryon kinetic energy has lower energy than the quark kinetic energy, whereas at larger distance, which corresponds to shorter distance to the neighbouring baryons,  the enhanced baryon-baryon repulsion over the quark-baryon repulsion favors quark excitation.  Such properties are exactly what we find from a constituent quark model analysis at sufficiently high baryon density.
However, as we will see, the initial shell-like momentum structure will involve the $d$-quark occupying the low momentum part connected by the baryon shell.
This is so because the $d$-quark excitation  in the neutron will leave the attractive ($ud$)-diquark intact and hence cost least amount of excitation energy.

The paper is organized as follows. In section II, we introduce the quarkyonic-like matter configuration.  In section III, we analyze the excitation modes of the baryon and quarks in the presence of a neighbouring nucleon.   In section IV, we find the lowest modes at a given nucleon density by extremization of the energies.  In the last section, we summarize our result and give prospects for further studies.

\section{Quarkyonic configuration}

In the phenomenological model of quarkyonic matter, one assumes that the nuclear matter undergoes a
 second-order or crossover phase transition at some moderate density regime since the first-order phase transition would not clearly appear due to the quark-hadron duality manifested in quarkyonic matter~\cite{McLerran:2007qj}.
Adopting these assumptions,
the phase space of fermions at these densities
is composed of two modes.  In the low momentum region, quasi-free
quark Fermi sea appears smoothly near the origin of the phase space.
Above the quasi-free quark modes, due to the presumed Pauli blocking effect, the confined quarks take the higher momentum region of the phase space through a shell-like baryon distribution in the phase space.  In such configuration,  the number density of each fermion can be counted from the phase space distribution:
\begin{align}
 n_{b_i}  &\equiv n_s \int^{\left[ k_{F} + \Delta \right]_{b_i}}_{k_{F}^{b_i}}\frac{d^3 k}{(2\pi)^3},\label{nb1}\\
 n_{\tilde{Q}_{l}}&\equiv  2 \int^{k_F^{Q_{l}}}_{0} \frac{d^3 k}{(2\pi)^3},\label{nb2}\\
 n_B &= \sum_i n_{b_i}+ \sum_l n_{\tilde{Q}_{l}}\label{nb3},
\end{align}
where Eq.~\eqref{nb1} and Eq.~\eqref{nb2} are  the contributions to the  baryon density coming from the baryon($b$) with flavor $i$ and quark($\tilde{Q}$) with flavor ${l}$, respectively.
The tilde in Eq.~\eqref{nb2} denotes the baryon density calculated in terms of the quark density so that the degeneracy of $N_c$ is compensated by the quark quantum number which is a factor of $1/N_c$ compared to that of the baryon. $n_s$ denotes the spin degeneracy.
Eq.~\eqref{nb3} is the total baryon density composed of quark part and baryon part.
Through Eq.~\eqref{nb1} the upper boundary of the baryon distribution is defined as $\left[ k_{F} + \Delta \right]_{b_i}=\left(({6\pi^2}/{  n_s})  n_{b_i} + {k_{F}^{b_i}}^3 \right)^{\frac{1}{3}}$. The lower boundary of the baryon distribution $k_{F}^{b_i}$ can be defined differently depending on the model assumption on the quark Fermi sea. For example, if one considers the isospin symmetric quasi-free quark sea, $k_{F}^{b}=N_c k_{F}^{\tilde{Q}}$.

For a given total baryon number density, the transition to the quarkyonic-like configuration occurs if the nontrivial minimum of the free energy (where $n_{\tilde{Q}}\neq 0$) is obtained. The free energy density at $T\rightarrow0$ can be written as
\begin{align}
f &=  \epsilon -Ts \nonumber\\
&\rightarrow \epsilon = n_s \sum_i   \int^{\left[ k_{F} + \Delta \right]_{b_i}}_{k_{F}^{b_i}}\frac{d^3 k}{(2\pi)^3} E_{b_i}(k) \nonumber\\
&\qquad \quad~+2\sum_{l}    \int^{k_F^{Q_{l}}}_{0} \frac{d^3 k}{(2\pi)^3} E_{Q_l}(k), \label{q-b-energy}
\end{align}
where $E_{b_i}(k)$ and $E_{Q_l}(k)$ represent the single baryon and quark energy at each phase cell, respectively. Whether the assumed two modes are the preferred state can be assessed  by extremizing  the free energy for a fixed density.
For the isospin symmetric dense matter, the extremization condition at a given total baryon density $n_B$ can be found as
\begin{align}
d\epsilon &= \frac{\partial \epsilon}{\partial n_{N}} d n_N +  \frac{\partial \epsilon}{\partial n_{\tilde{Q}}} d n_{n_{\tilde{Q}}},\nonumber\\
&= \left( \frac{\partial \epsilon}{\partial n_{N}}- \frac{\partial \epsilon}{\partial n_{\tilde{Q}}} \right) d n_{N} =0,
\end{align}
where $n_N$ denotes the nucleon density. One can find the condition $\mu_N=\mu_{\tilde{Q}}$ for the nontrivial minimum ($n_{\tilde{Q}}\neq0$). If one assumes a diverging repulsive potential at short interdistance of baryon, this relation is satisfied around the hard-core density~\cite{Jeong:2019lhv, Sen:2020peq, Duarte:2020xsp, Duarte:2020kvi, Sen:2020qcd}.
In the low nuclear density regime as well as at the normal nuclear matter, the free energy should have the trivial minimum with $n_{\tilde{Q}}=0$.

In this work, we will study the possible appearance of the $d$-quarks in dense neutron rich matter as the dominating long-range modes in the configuration space (low momentum modes in phase space). This is accomplished by calculating the quark and baryon energy relevant for Eq.~\eqref{q-b-energy} using within the constituent quark model. While the energies will be expressed in terms of the position and momentum coordinates, we can express them into momentum coordinates since the lowest energy mode of the fermions satisfy the minimum uncertainty
\begin{align}
r =\frac{1}{k},\label{rp}
\end{align}
which is particularly suited for the trial wave function of a gaussian form as used in Ref.~\cite{Park:2019bsz} to describe the ground state of a multiquark configuration.

\section{Excitation energies}

\begin{figure}
\includegraphics[width=0.45\textwidth]{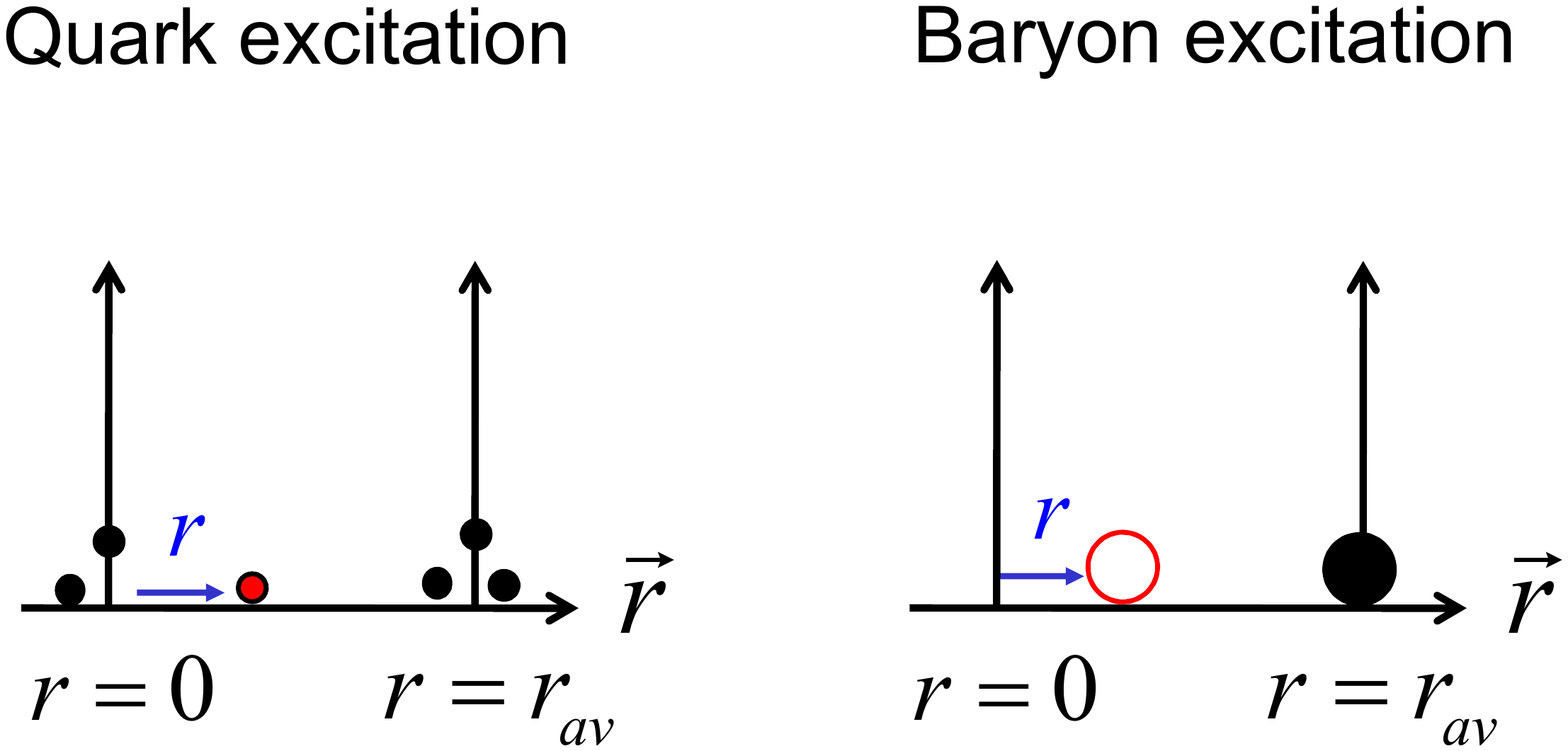}
\caption{Left: Excitation of the quark.  Right: Excitation of the baryon. }
\label{twomodes}
\end{figure}

Let us consider two nucleons separated by a distance $r_{av}$, which we will assume is the inter-baryon distance at finite baryon density $n\sim 1/r_{av}^3$.
We will  consider the lowest possible excitation modes in the quark and baryon basis.
 For that purpose, we first consider the excitation of the baryon  as a function of $r$ from its equilibrium position as shown in the right figure of Fig. \ref{twomodes}.  Second, we explore the interaction of the lowest quark mode as shown in the left figure.  All remaining quark and baryon configurations and their relative interaction energies are assumed to be fixed.

\subsection{Excitation of the baryon mode}

Here we first discuss the right figure of Fig. \ref{twomodes}.
 While the long and intermediate nuclear attraction is important in providing
nuclear saturation, here we will neglect these interactions and concentrate on the nuclear repulsion with the neighbouring nucleon.
This is so because the quark mode only appears near the neighboring baryon (small $|r-r_{av}|$) because the short distance quark-baryon repulsion is smaller  compared to the baryon-baryon repulsion at same separation.
Therefore, any  intermediate attraction in the baryon potential will only
strengthen the baryonic mode at that position.
Hence, the dominant baryon energy is just the repulsion due to the baryon at $r=r_{av}$.

To parametrize the relevant interaction within a single model, we will use the following quark model Hamiltonian, which was recently shown to explain the lattice data for nuclear repulsion at short distance~\cite{Park:2019bsz}.
\begin{eqnarray}
H=\sum_{i=1}^{N}(m_{i}+\frac{\textbf{p}^2_i}{2m_i})
 -\frac{3}{16}\sum_{i<j}^{N}(V^{C}_{ij}+V^{CS}_{ij}),\label{Hamiltonian}
\end{eqnarray}
where $N$ is the total number of constituent quarks and  $m_i$'s are the constituent quark masses.
The spin-independent (spin-dependent)  color interaction denoted as
$V^{C}_{ij}$ ($V^{CS}_{ij}$) is given by  \cite{Bhaduri:1981pn,Park:2016mez}.
\begin{eqnarray}
V^{C}_{ij}=   \lambda^c_i\lambda^c_j \left( - \frac{\kappa}{r_{ij}}+\frac{r_{ij}}{a_0}-D \right) ,
\end{eqnarray}
\begin{eqnarray}
 V^{CS}_{ij}
=\frac{{\kappa}^{\prime}}{m_im_j r_{0ij}^2 } \frac{1}{r_{ij}}e^{-(r_{ij}/r_{0ij})^2} \lambda^c_i\lambda^c_j \ {\sigma}_i\cdot{\sigma}_j , \label{vcs}
\end{eqnarray}
where
$\lambda^c_i$ are the Gell-Mann matrices of the $i$'th quark for the color SU(3) and $\sigma_i$ are the Pauli matrices for the spin SU(2). Here, $r_{ij}$ is the distance between quarks, while $r_{0ij}$ is chosen to depend on
the constituent quark masses as
\begin{eqnarray}
r_{0ij}=  (\alpha+\beta \mu_{ij})^{-1}. \label{mass-cs}
\end{eqnarray}
with $\mu_{ij}=  {m_im_j}/({m_i+m_j})$ being the
reduced mass. The parameters are given in Table \ref{parameters}.

\begin{center}
\begin{table}[t]
  \begin{tabular}{|c|c|c|c|c|c|c|c|c|c|}
     \hline
    $\kappa$  & $\kappa '$ & $a_0$  & $D$ & $\alpha$ &$\beta$ & $m_{u,d}$    \\
    \hline \hline
    0.59  & 0.5 & 5.386 & 0.96  &  2.1    & 0.552 & 0.343  \\
          &     & $ \mbox{GeV}^{-2}$ & $\mbox{GeV}$ & $\mbox{fm}^{-1}$  & & $\mbox{GeV}$ \\
    \hline
  \end{tabular}
   \caption{The  parameters of CQM fitted to baryon masses \cite{Park:2016mez}. }
  \label{parameters}
\end{table}
\end{center}

\begin{figure}
  \includegraphics[width=0.4\textwidth]{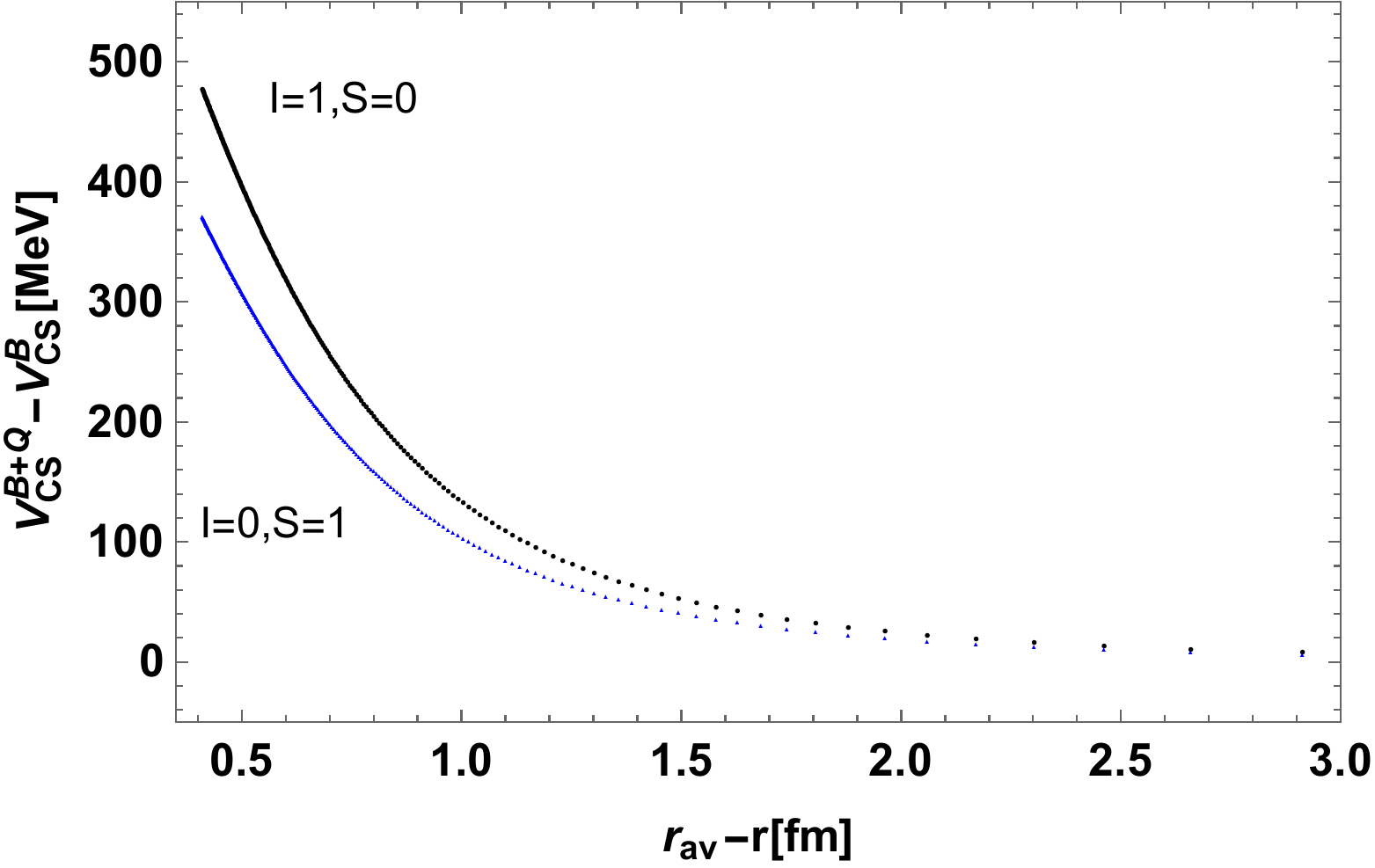}
  \caption{Nucleon-Nucleon repulsion obtained from the hyperfine potential in Eq. \eqref{Hamiltonian}.}
\label{Fig_CS2}
\end{figure}

There are two possible states of two nucleons satisfying the Pauli principle: $(I,S)=(1,0),(0,1)$. Fig.~\ref{Fig_CS2}  shows the nucleon-nucleon repulsions calculated from the constituent quark model using  Eq. (\ref{Hamiltonian}) for the two channels. We calculate the nucleon-nucleon repulsions using a simple Gaussian function as a spatial part of the multiquark wave function.
Including the kinetic term and neglecting all other nuclear interaction, we then find the nonrelativistic Hamiltonian of a baryon for $I=1,S=0$ to be
\begin{eqnarray}
E_B^{I=1,S=0} & = & m_B+\frac{k^2}{2m_{B}}
 \nonumber\\ &&
+ \frac{a_B }{|r-r_{av}|}
e^{-\frac{|r-r_{av}|^2}{b_B^2 }} - \frac{a_B }{|r_{av}|}e^{-\frac{|r_{av}|^2}{b_B^2}},
 \label{e-b}
\end{eqnarray}
where   $m_B$ is the nucleon mass.
The fit to the plot in Fig.~\ref{Fig_CS2} which is calculated using a constituent quark model gives $a_B=0.218$ GeV$\cdot$fm and $b_B=1.474$ fm. The potential part of Eq.~\eqref{e-b} comes from the hyperfine interaction in Eq.~\eqref{vcs}, and has been normalized to vanish at $r=0$ as the baryon mode and quark mode should describe the same state when $r=0$. The quark potential will be normalized in the same way.  Since the difference between $I=0$ and $I=1$ channels is small, we will take $E_B^{I=1,S=0}$ as the baryon mode energy  $E_B$ for both channels.

\subsection{Excitation of the lowest energy quark mode}

Here we discuss the left figure of Fig.~\ref{twomodes}.
For that purpose, consider pulling away one quark from the remaining two quarks.  For a neutron composed of $(udd)$ quarks, it could be either a $u$ quark or a $d$ quark.

\begin{figure}
  \includegraphics[width=0.4\textwidth]{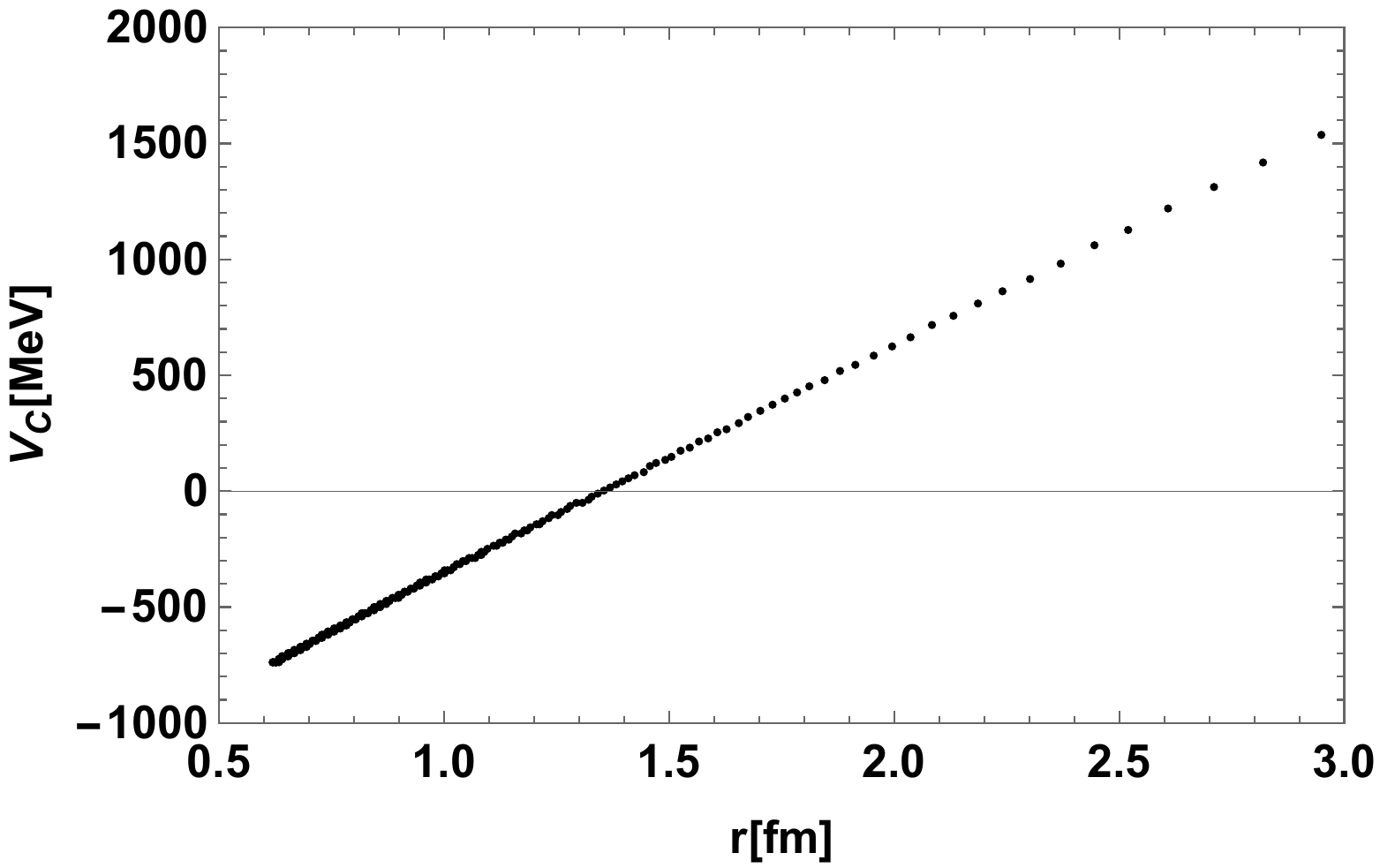}
  \caption{Confinement potential of a $d$ quark.}
\label{Fig_C1}
\end{figure}

We note that the difference in pulling out a $d$ or $u$ quark lies in the color-spin interaction.  This can be intuitively understood by considering the strength of the color-spin factor among quarks.
\begin{align}
  {\cal X} & \equiv
  -\sum_{i< j}^N \langle \lambda_i^c \lambda_j^c\ \sigma_i \cdot \sigma_j \rangle  \nonumber \\
  &
  = N(N-10)+\frac{4}{3}S(S+1)+ 4C_{\rm F}+2C_{\rm C}.
\label{color-spin-formula}
\end{align}
Here   $C_{\rm F}$ and $C_{\rm C}$ are
the first kind of Casimir operators of flavor and color of SU(3) for the $N$-quark system, respectively \cite{Aerts:1977rw}.
This color-spin factor is $-8$ and $+8$ for the nucleon and delta, respectively, and is responsible for their mass difference. As discussed in Ref.~\cite{Park:2019bsz}, the color-spin interaction based on this factor can explain the recent lattice calculation for baryon-baryon interaction at short distance.
For the nucleon, the factor comes from the attraction among three quarks.  However, the ($I=0,S=0$) color antitriplet diquark $(ud)$, which is often called the most attractive diquark and comprises the dominant component of the $ud$ quark component inside a nucleon, also has $ {\cal X}=-8$ .  Therefore, while the pulling away a $u$ quark from the neutron will cost color-spin energy, the excitation of a $d$ quark will not cost any color-spin energy as the attractive $(ud)$ diquark remains intact.
Furthermore, the confining potential for $u,d$ quarks is the same.  Therefore, $d$ quark excitation mode constitutes the lowest quark excitation mode  inside a neutron.  As for the effect of confining potential, we represent the result of $d$ quark in Fig. \ref{Fig_C1}.

To calculate the repulsive potential between the excited $d$ quark and the neighbouring nucleon, we consider the four-quark states.
For the four-quark state, there are four possible states  satisfying the Pauli principle: $(I,S)=(0,0),(0,1),(1,0),(1,1)$. For a neutron and a $d$ quark, the lowest energy state is $I=1,S=0$.

\begin{widetext}
The quark model calculation for the $d$ quark potential from the remaining two quarks and the neighbouring nucleon  can be parametrized as
\begin{eqnarray}
E_Q & = & m_B-m_D + \frac{k^2}{2m_Q} +\sigma_n \left( r H(r_{\text{max}}-r) + r_{\text{max}}H(r-r_{\text{max}}) \right) +  R_n \left( \frac{a_Q }{|r-r_{av}|}
e^{-\frac{|r-r_{av}|}{b_Q }} - \frac{a_Q }{|r_{av}|}e^{-\frac{|r_{av}|}{b_Q }} \right),
\label{e-q1}
\end{eqnarray}
\end{widetext}
where $m_D$ is the mass of remaining $ud$ diquark, $H(r)$ is the Heaviside step function, and the parameters are  $m_Q=0.343~\textrm{GeV}$, $a_Q = 0.2~\textrm{GeV}\cdot\textrm{fm}$ and $b_Q = 0.754~\textrm{fm}$.
The fourth term is from the confinement potential between the $d$-quark and remaining two quarks, while the sixth term is the quark-nucleon repulsion. The confinement potential becomes constant when the string breaks at $r=r_{\text{max}}$. This parameter is determined so that the sum of confinement potential and kinetic energy becomes the mass of the quark-antiquark pair.
Also, we assume that the string tension decreases linearly in nuclear medium with the same fractional change as the gluon condensate~\cite{Hatsuda:1991ez}:
\begin{eqnarray}
\sigma_n & = & \sigma \left(1-0.05 \frac{n}{n_0}\right)
\end{eqnarray}
where $n_0=0.16~\textrm{fm}^{-3}$ is the nuclear saturation density and $\sigma = 0.962 $~GeV/fm.

\begin{figure}
  \includegraphics[width=0.4\textwidth]{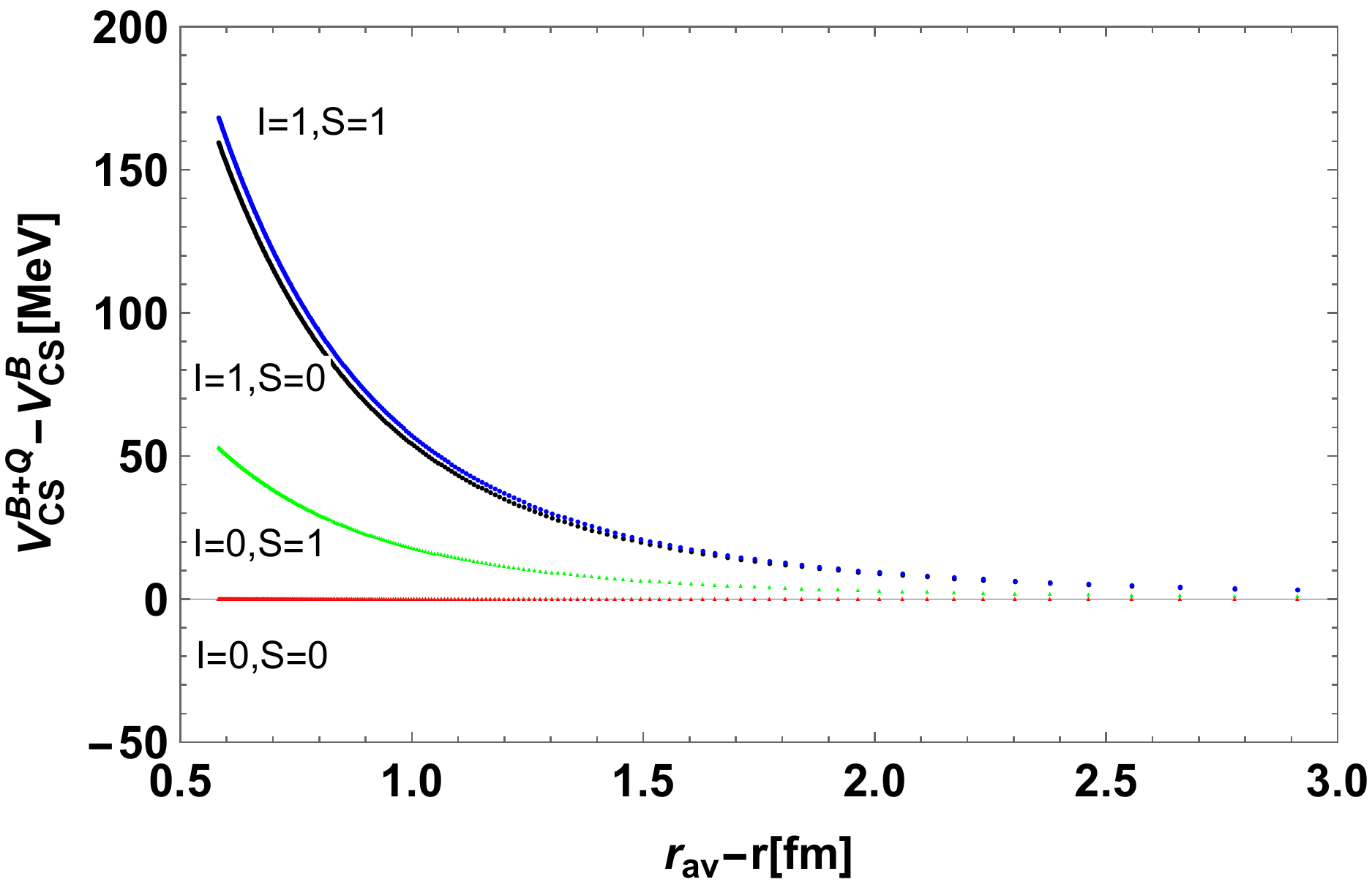}
  \caption{Nucleon-$d$ quark repulsion obtained from the hyperfine potential in Eq. \eqref{Hamiltonian}.}
\label{Fig_CS1}
\end{figure}

The parameters in the quark-nucleon repulsion is
extracted from the $I=1$ and $S=0$ four-quark configuration. We represent the repulsion from the hyperfine potential for all possible configurations in Fig. \ref{Fig_CS1}.
$R_n$ accounts for the contribution from the $I=0$ four-quark configuration.
Hence, if we set $R_n=1$, then $E_Q=E_Q^{I=1,S=0}$. However, as the nuclear matter in a neutron star also contains protons, we have to consider $I=0$ states as well as $I=1$. As the four-quark repulsion vanishes in the $I=S=0$ channel,  we multiply the repulsion terms by a  ratio factor, which we take to be $R_n=0.7$ in this work: smaller values will lead to slightly smaller density for quark modes to appear.  The last term in Eq.~\eqref{e-q1} is added so that the repulsive term vanishes  when $r$ goes to zero. One should note that the repulsion of quark-baryon shows slightly different exponential form compared to the baryon-baryon repulsion.

Additionally, it should be noted that Eq. \eqref{e-q1} is valid only when
$r  <  r_{av}$ and $k<k_{\text{Fermi}}$,  so that the valid range in momentum space becomes
\begin{eqnarray}
\frac{1}{r_{av}} <k <k_{\text{Fermi}}. \label{range}
\end{eqnarray}

Now, it is very important that
\begin{eqnarray}
a_B & > & a_Q, \nonumber \\
m_Q & < & m_B.
\end{eqnarray}
These two conditions eventually lead the quark excitation mode to be dominant at smaller momentum while the baryon excitation mode becomes important at larger momentum.

\section{Extremization}

In the previous section, we have demonstrated that the colored $d$ quark and $ud$ diquark configuration rather than the neutron can be the energetically favourable state if one consider the long-range separation of quarks from the equilibrium position in dense circumstance. Since the lowest energy mode of constituent quarks satisfy the minimal uncertainty~\eqref{rp}, the separation distance scale of the constituent quark can be approximated in terms of the momentum fluctuation scale of the quarks. By taking the two-body potentials~\eqref{e-b} and \eqref{e-q1} as centrifugal mean-field potentials for the quasi-fermions in dense neutron matter, the onset condition for the $d+ud$ configuration can be examined in an analogous way to the original approach~\cite{McLerran:2007qj, McLerran:2018hbz}.

Assuming the Pauli exclusion principle between the $d$ quark and the confined quarks in its neighbouring neutrons, the extremized configuration can be found in terms of $E_Q$ and $E_B$ following Ref.~\cite{Jeong:2019lhv, Sen:2020peq, Duarte:2020xsp, Duarte:2020kvi, Sen:2020qcd}. If the neutron is turned into the $ud$ diquark and $d$ quark, Eqs.~\eqref{nb1}-\eqref{nb3} can be revised as follows:
\begin{align}
 n_{n}  &=2 \int^{\left[ k_{F} + \Delta \right]_{n}}_{k_{F}^{n}}\frac{d^3 k}{(2\pi)^3},\label{relation-1}\\
 n_{d}&= n_{ud}= 2 \int^{k_F^{d}}_{0} \frac{d^3 k}{(2\pi)^3},\label{relation-2}\\
 n_B&=  n_{n}+ \frac{n_{d}+2n_{ud}}{3}\label{relation-3},
\end{align}
where the subscripts $n,d,ud,B$ respectively denote the neutron,  $d$-quark with a color in the triplet,  $(ud)$-diquark with a color in the anti-triplet  and  baryon. The outer boundary of the neutron distribution is defined as
\begin{align}
\left[k_{F} + \Delta \right]_{n} = \left( 3\pi^2 (n_{n} + n_{d})\right)^{\frac{1}{3}} \label{relation-4},
\end{align}
 where the color of  $ud$ diquark and $d$ quark are  correlated. The corresponding energy density can be written as
\begin{align}
\epsilon  &=2 \int^{\left[ k_{F} + \Delta \right]_{n}}_{k_{F}^{n}}\frac{d^3 k}{(2\pi)^3} E_{n}(k) + 2\int^{k_F^{d}}_{0} \frac{d^3 k}{(2\pi)^3} E_{d}(k)\nonumber\\
&\quad+m_D n_{ud},\label{relation-5}
\end{align}
where $E_n$ and $E_d$ are obtained from Eq.~\eqref{e-b} and Eq.~\eqref{e-q1}, respectively. $m_D$ represents the remaining mass related to the diquark energy.

The extremum appears when $d\epsilon=0$:
\begin{align}
d\varepsilon &= \frac{\partial \varepsilon}{\partial n_{n}} d n_{n} +  \frac{\partial \varepsilon}{\partial n_{d}} d n_{n_{d}}+  \frac{\partial \varepsilon}{\partial n_{ud}} d n_{n_{ud}}\nonumber\\
&= \left( \frac{\partial \varepsilon}{\partial n_{n}}- \frac{\partial \varepsilon}{\partial n_{d}} - \frac{\partial \varepsilon}{\partial n_{ud}}\right) d n_{n} =0,\label{relation-6}
\end{align}
where $d n_{B}=dn_{n}+dn_{d}+dn_{ud}=0$. The chemical potential can be obtained as follows:
\begin{align}
 \mu_{n} &=\frac{\partial \varepsilon}{\partial n_{n}}\nonumber\\
 &=\frac{  \left[k_F+\Delta\right]_{n}^2}{\pi^2} E_n(\left[k_F+\Delta\right]_{n}) \frac{\partial  \left[k_F+\Delta\right]_{n}}{\partial n_{n} }\nonumber\\
 &=E_n(\left[k_F+\Delta\right]_{n}),\label{relation-7}\\
\mu_{d} + \mu_{ud}&= \frac{\partial \varepsilon}{\partial n_{d}}+\frac{\partial \varepsilon}{\partial n_{ud}}\nonumber\\
& = \frac{ \left[k_F+\Delta\right]_{n}^2}{\pi^2} E_n(\left[k_F+\Delta\right]_{n}) \frac{\partial \left[k_F+\Delta\right]_{n} }{\partial  n_{\tilde{d}}} \nonumber\\
&\quad - \frac{  { k^{n}_F }^2}{\pi^2} E_n( k^{n}_F)  \frac{\partial  k^{n}_F  }{\partial n_{\tilde{d}} } + N_c E_d( k^{d}_F) \nonumber\\
&=E_n(\left[k_F+\Delta\right]_{n})  -E_n( k^{n}_F) \nonumber\\
&\quad+ E_d( k^{d}_F)+m_D,\label{relation-8}
\end{align}
where  $k^{n}_F= k^d_F$ is understood and the partial derivatives can be found as
\begin{align}
\frac{\partial  \left[k_F+\Delta\right]_{n}}{\partial n_{n} }&=\frac{ \pi^2 }{ \left[k_F+\Delta\right]_{n}^2},\label{relation-9}\\
 \frac{\partial \left[k_F+\Delta\right]_{n} }{\partial  n_{d}}&=\frac{ \pi^2 }{ \left[k_F+\Delta\right]_{n}^2},\label{relation-10}\\
  \frac{\partial  k^{n}_F }{\partial n_{d} } &=\frac{ \pi^2 }{{ k^{n}_F}^2}\label{relation-11}.
\end{align}
Substituting Eqs.~\eqref{relation-7}-\eqref{relation-11} into
Eq.~\eqref{relation-6}, the extremum appears when the following energy difference vanishes.
\begin{eqnarray}
\Delta E =    E_d( k^{d}_F)+m_D-E_n( k^{n}_F=k^d_F).  \label{condition}
\end{eqnarray}
 Since only  a colored $d$ quark together with the static $ud$ diquark appears in this configuration, the disconnected shell-like momentum distribution, which causes a stiff evolution of EoS, does not appear yet. The lower phase of the neutron Fermi sea is dissociated into $d$ quark and $ud$ diquark, which can be related to the modified quark phase measure introduced in Ref.~\cite{Jeong:2019lhv} where the Fermi momentum of the quasi-quark sea is reduced by IR cutoff around the onset of the quasi-quark sea. The genuine quarkyonic-like configuration might appear at higher densities, where multi-neutron potential and deconfinement become important.

\begin{widetext}
\begin{center}
\begin{figure}
  \includegraphics[width=0.8\textwidth]{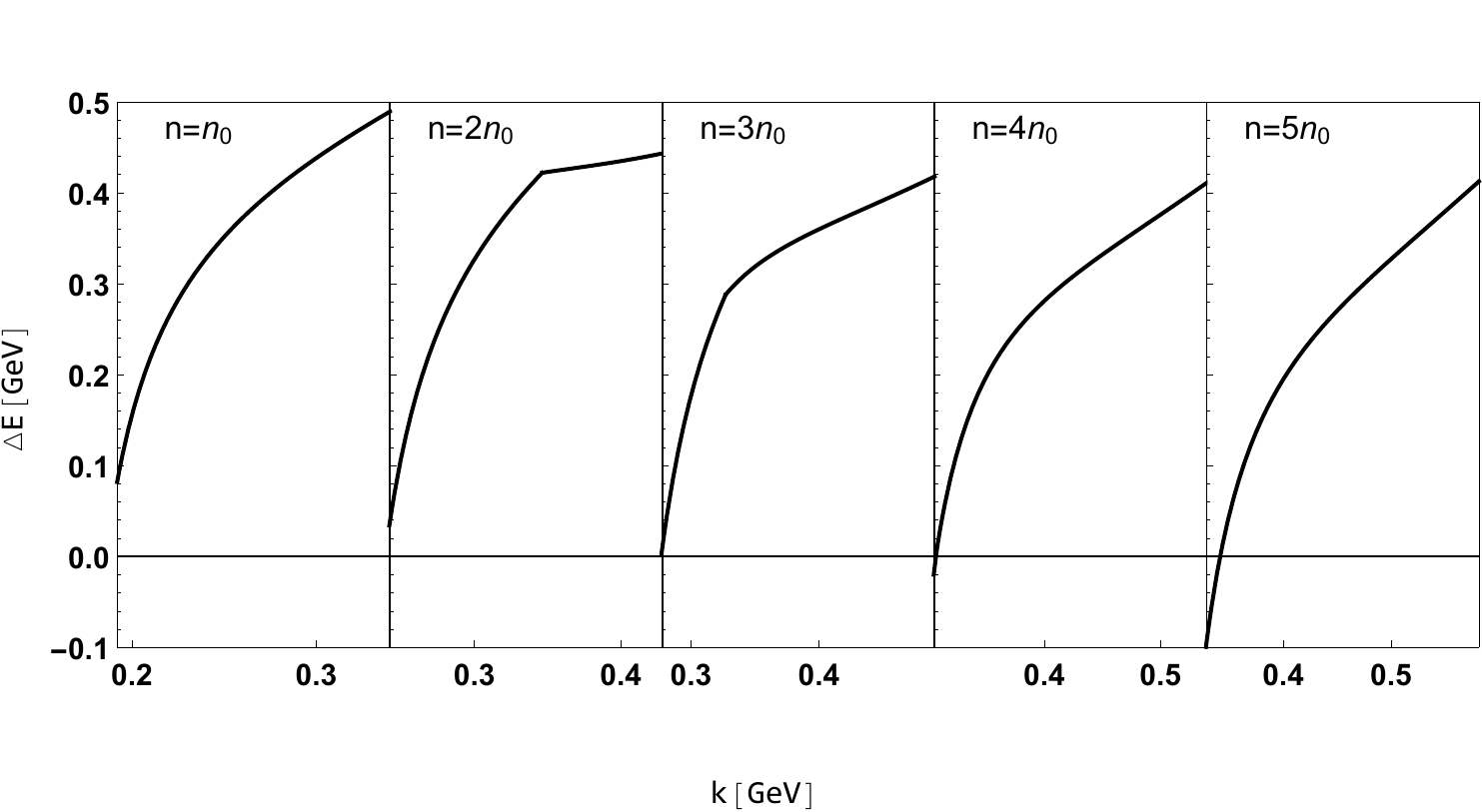}
  \caption{$\Delta E$ where $R_{n}=0.7$. Here, $n_0$ is a normal nuclear density.}
\label{Fig_ED}
\end{figure}
\end{center}
\end{widetext}

Fig.~\ref{Fig_ED} shows the energy difference represented in Eq.~\eqref{condition} within the momentum range in Eq.~\eqref{range} for selected nucleon densities.
 We can check that, at high nuclear density ($n > 4n_0$), the energy of the $d$ quark excitation mode is smaller than the baryon excitation mode at low momentum.  Such behavior  is consistent with the formation of the quarkyonic-like momentum   space structure suggested in Refs.~\cite{McLerran:2018hbz, Jeong:2019lhv}.
 Although this analysis is based on semi classical approximations using the expectation value of the effective interaction, which is   not based on the full quantum process, the complete algebraic relation implies that the baryon quantum number could be understood in either  baryon or quark according to the probing scale to the matter~\cite{McLerran:2007qj}.

\section{Summary}

The quarkyonic matter configuration is originally suggested as the quark-baryon coexistence phase where the Fermi-Dirac statistics is well defined between all the confined and quasi-free quarks through the quark-hadron duality. Thus, in the phenomenological sense, the onset of the quarkyonic matter could be understood as the transition moment where the quasi-free quarks and multiquark substructure appear smoothly in the ground state of the dense matter.

In this work, we analyzed to what extent the quarkyonic modes appear in the phase space of baryons as one increases the density. For that purpose, we first considered the baryon excitation mode as a function of $r$ from its equilibrium position using a constituent quark model. For the quark excitation mode, since pulling away a $d$ quark will not cost any color-spin energy as the attractive $(ud)$ diquark remain intact, we calculated the $d$ quark excitation mode from a neutron in a similar way as a baryon excitation. To solve the excitation mode in the phase space, we assumed that $r={1}/{k}$. We find that at high nuclear density ($n >4n_0$), the energy of a $d$ quark excitation mode is smaller than a baryon excitation mode starting at lower momentum.
The results show that the repulsion between nucleons at short distance leads to
quarkyonic-like matter but that the initial quark excitation mode will involve a single quark that leaves the attractive diquark intact.

\begin{widetext}
\begin{center}
\begin{figure}
  \includegraphics[width=0.8\textwidth]{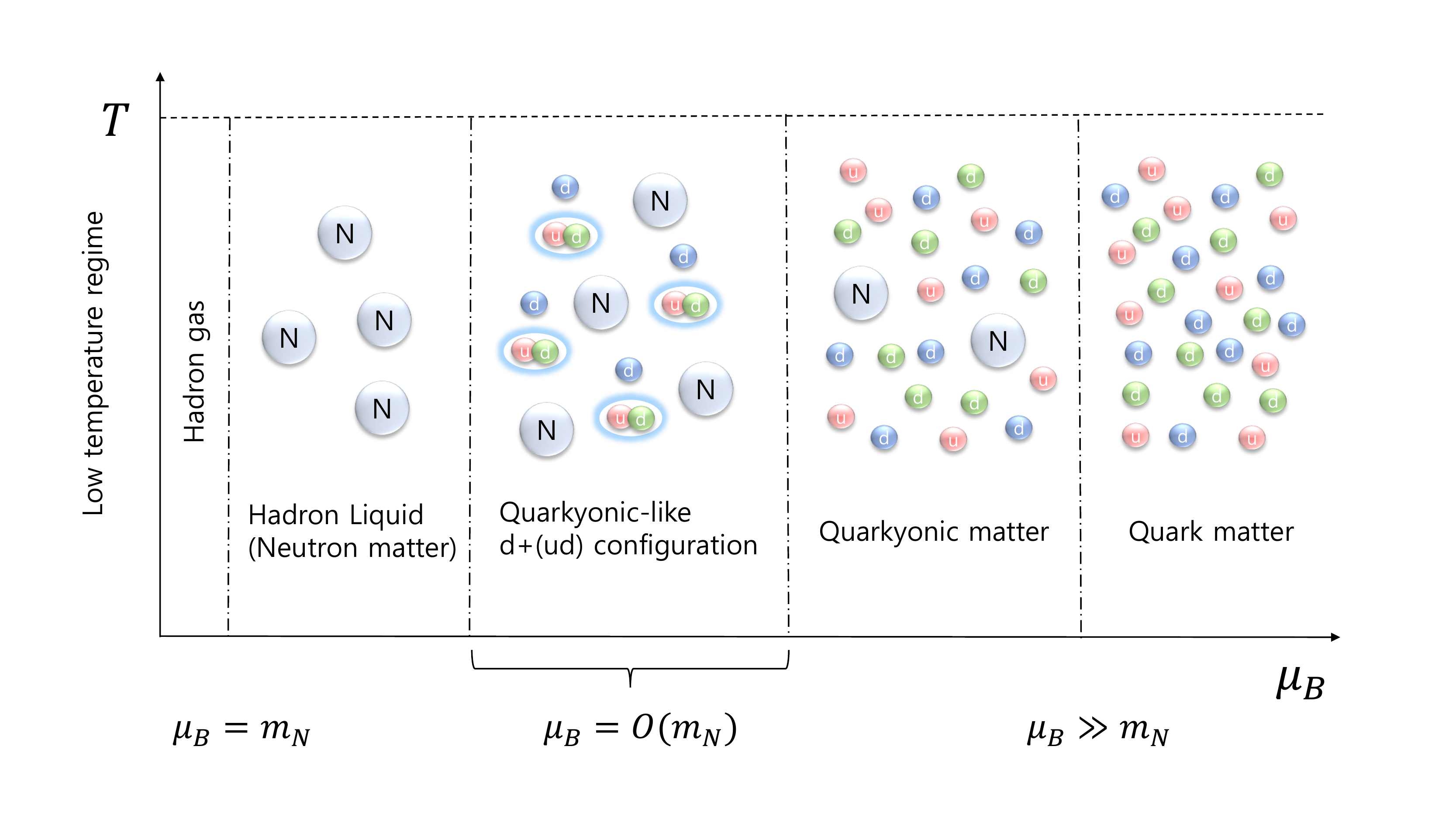}
  \caption{A schematic phase diagram anticipated from the constituent quark model. This anticipation follows the conclusion of Ref.~\cite{McLerran:2007qj}}
\label{pd}
\end{figure}
\end{center}
\end{widetext}

Here we assumed the color-spin interaction of the consitituent quark as the dominant contribution to the dense matter dynamics follwing Ref.~\cite{Park:2019bsz} and compared the relative energy of each possible mode.To obtain the quantitatively realistic nuclear bulk properties such as the compressibility and the EoS, one needs to consider density dependence of  the background mean-field potential for the quasi-nucleons at high densities.  If the sum of quark-meson and diquark-meson interactions do not differ much from the neutron-meson interaction, the pressure coming from the meson exchange potential should not increase drastically by the formation of the $d$ quark and $ud$ diquark configuration. However, when having the individual quark modes is preferred than keeping the diquark modes, the shell-like phase space distribution becomes evident and the pressure should increase stiffly with the anomalous sound velocity regardless of the quark-meson interaction (Fig.~\ref{pd}).

It should be noted that this analysis is based on simple two-body interactions
where the  interaction from the neighbouring nucleons are taken into account through a average separation distance.  In a more realistic case, we need to consider the three-dimensional configuration and take into account higher-body interactions. Additionally, it is necessary to consider not only the $d$ quark but also the excitation of all three quarks with varying confinement effects. We will discuss the onset condition for the genuine quarkonyic configuration in the next paper. The discussion will provide a theoretical support for the neutron star based on quarkyonic matter concept  where the incompressible core and the soft outer parts appear naturally.

\acknowledgements

This work was supported by Samsung Science and Technology Foundation under Project Number SSTF-BA1901-04. K.S.J acknowledge the support of the Simons Foundation under the Multifarious Minds Program Grant No. 557037. The work of K.S.J. was supported by the U.S. DOE under Grant No. DE-FG02-00ER41132. The work of A.P. was supported by the Korea National Research Foundation under the grant number 2021R1I1A1A01043019.


\begin{thebibliography}{99}

\bibitem{Demorest:2010bx}
P.~Demorest, T.~Pennucci, S.~Ransom, M.~Roberts and J.~Hessels,
Nature \textbf{467}, 1081-1083 (2010).


\bibitem{Antoniadis:2013pzd}
J.~Antoniadis {\it et al.}
Science \textbf{340}, 6131 (2013)


\bibitem{Cromartie:2019kug}
H.~T.~Cromartie {\it et al.}
Nature Astron. \textbf{4}, no.1, 72-76 (2019)

\bibitem{TheLIGOScientific:2017qsa}
  B.~P.~Abbott {\it et al.} [LIGO Scientific and Virgo Collaborations],
  Phys.\ Rev.\ Lett.\  {\bf 119}, no. 16, 161101 (2017)


\bibitem{Abbott:2018exr}
  B.~P.~Abbott {\it et al.} [LIGO Scientific and Virgo Collaborations],
  Phys.\ Rev.\ Lett.\  {\bf 121}, no. 16, 161101 (2018)

\bibitem{Annala:2017llu}
  E.~Annala, T.~Gorda, A.~Kurkela and A.~Vuorinen,
  Phys.\ Rev.\ Lett.\  {\bf 120}, no. 17, 172703 (2018)

\bibitem{Raithel:2018ncd}
  C.~Raithel, F.~Özel and D.~Psaltis,
  Astrophys.\ J.\  {\bf 857}, no. 2, L23 (2018)


\bibitem{Fujimoto:2019hxv}
Y.~Fujimoto, K.~Fukushima and K.~Murase,
Phys. Rev. D \textbf{101}, no.5, 054016 (2020)

\bibitem{Miller:2021qha}
M.~C.~Miller, F.~K.~Lamb, A.~J.~Dittmann, S.~Bogdanov, Z.~Arzoumanian, K.~C.~Gendreau, S.~Guillot, W.~C.~G.~Ho, J.~M.~Lattimer and M.~Loewenstein, \textit{et al.}
[arXiv:2105.06979 [astro-ph.HE]].

\bibitem{Riley:2021pdl}
T.~E.~Riley, A.~L.~Watts, P.~S.~Ray, S.~Bogdanov, S.~Guillot, S.~M.~Morsink, A.~V.~Bilous, Z.~Arzoumanian, D.~Choudhury and J.~S.~Deneva, \textit{et al.}
[arXiv:2105.06980 [astro-ph.HE]].



\bibitem{McLerran:2007qj}
  L.~McLerran and R.~D.~Pisarski,
  Nucl.\ Phys.\ A {\bf 796}, 83 (2007)




\bibitem{tHooft:1973alw}
  G.~'t Hooft,
  Nucl.\ Phys.\ B {\bf 72}, 461 (1974).


\bibitem{tHooft:1974pnl}
  G.~'t Hooft,
  Nucl.\ Phys.\ B {\bf 75}, 461 (1974).




\bibitem{McLerran:2018hbz}
  L.~McLerran and S.~Reddy,
  Phys.\ Rev.\ Lett.\  {\bf 122}, no. 12, 122701 (2019)



\bibitem{Jeong:2019lhv}
K.~S.~Jeong, L.~McLerran and S.~Sen,
Phys. Rev. C \textbf{101}, no.3, 035201 (2020)




\bibitem{Sen:2020peq}
S.~Sen and N.~C.~Warrington,
Nucl. Phys. A \textbf{1006}, 122059 (2021)



\bibitem{Duarte:2020xsp}
D.~C.~Duarte, S.~Hernandez-Ortiz and K.~S.~Jeong,
Phys. Rev. C \textbf{102}, no.2, 025203 (2020)



\bibitem{Duarte:2020kvi}
D.~C.~Duarte, S.~Hernandez-Ortiz and K.~S.~Jeong,
Phys. Rev. C \textbf{102}, no.6, 065202 (2020)



\bibitem{Sen:2020qcd}
S.~Sen and L.~Sivertsen,
[arXiv:2011.04681 [astro-ph.HE]].

\bibitem{Zhao:2020dvu}
T.~Zhao and J.~M.~Lattimer,
Phys. Rev. D \textbf{102}, no.2, 023021 (2020)


\bibitem{Margueron:2021dtx}
J.~Margueron, H.~Hansen, P.~Proust and G.~Chanfray,
[arXiv:2103.10209 [nucl-th]].


\bibitem{Bethe:1971xm}
H.~Bethe,
Ann. Rev. Nucl. Part. Sci. \textbf{21}, 93-244 (1971)


\bibitem{Lacombe:1980dr}
  M.~Lacombe, B.~Loiseau, J.~M.~Richard, R.~Vinh Mau, J.~Conte, P.~Pires and R.~de Tourreil,
  Phys.\ Rev.\ C {\bf 21}, 861 (1980).

\bibitem{Machleidt:1987hj}
  R.~Machleidt, K.~Holinde and C.~Elster,
  Phys.\ Rept.\  {\bf 149}, 1 (1987).

\bibitem{Oka:1981rj}
  M.~Oka and K.~Yazaki,
  Prog.\ Theor.\ Phys.\  {\bf 66}, 572 (1981).

\bibitem{Shimizu:1989ye}
  K.~Shimizu,
  Rept.\ Prog.\ Phys.\  {\bf 52}, 1 (1989).

\bibitem{Weinberg:1990rz}
  S.~Weinberg,
  Phys.\ Lett.\ B {\bf 251}, 288 (1990).

\bibitem{Ordonez:1993tn}
  C.~Ordonez, L.~Ray and U.~van Kolck,
  Phys.\ Rev.\ Lett.\  {\bf 72}, 1982 (1994).

\bibitem{Bedaque:2002mn}
  P.~F.~Bedaque and U.~van Kolck,
  Ann.\ Rev.\ Nucl.\ Part.\ Sci.\  {\bf 52}, 339 (2002)

\bibitem{Polinder:2007mp}
  H.~Polinder, J.~Haidenbauer and U.-G.~Meissner,
  Phys.\ Lett.\ B {\bf 653}, 29 (2007)

\bibitem{Inoue:2010hs}
  T.~Inoue {\it et al.} [HAL QCD Collaboration],
  Prog.\ Theor.\ Phys.\  {\bf 124}, 591 (2010)

\bibitem{Ishii:2006ec}
  N.~Ishii, S.~Aoki and T.~Hatsuda,
  Phys.\ Rev.\ Lett.\  {\bf 99}, 022001 (2007)

\bibitem{Park:2019bsz}
A.~Park, S.~H.~Lee, T.~Inoue and T.~Hatsuda,
Eur. Phys. J. A \textbf{56}, no.3, 93 (2020)


\bibitem{Bhaduri:1981pn}
  R.~K.~Bhaduri, L.~E.~Cohler and Y.~Nogami,
  Nuovo Cim.\ A {\bf 65}, 376 (1981).

\bibitem{Park:2016mez}
  A.~Park, W.~Park and S.~H.~Lee,
  Phys.\ Rev.\ D {\bf 94},  054027 (2016)

\bibitem{Aerts:1977rw}
  A.~T.~M.~Aerts, P.~J.~G.~Mulders and J.~J.~de Swart,
  Phys.\ Rev.\ D {\bf 17}, 260 (1978).

\bibitem{Hatsuda:1991ez}
T.~Hatsuda and S.~H.~Lee,
Phys. Rev. C \textbf{46}, no.1, R34 (1992)





















	








\end{thebibliography}
\end{document}